# Sensitivity of Water Dynamics to Biologically Significant Surfaces of Monomeric Insulin: Role of Topology and Electrostatic Interactions

## Kushal Bagchi[†] and Susmita Roy[*‡]

†St Joseph's College for Arts and Science, Bangalore 560027, India.

‡SSCU, Indian Institute of Science, Bangalore 560012, India.

## *Abstract*

In addition to the biologically active monomer of the protein Insulin circulating in human blood, the molecule also exists in dimeric and hexameric forms that are used as storage. The Insulin monomer contains two distinct surfaces, namely the dimer forming surface (DFS) and the hexamer forming surface (HFS) that are specifically designed to facilitate the formation of the dimer and the hexamer, respectively. In order to characterize the structural and dynamical behaviour of interfacial water molecules near these two surfaces (DFS and HFS), we performed atomistic molecular dynamics simulations of Insulin with explicit water. Dynamical characterization reveals that the structural relaxation of the hydrogen bonds formed between the residues of DFS and the interfacial water molecules is faster than those formed between water and that of the HFS. Furthermore, the residence times of water molecules in the protein hydration layer for both the DFS and HFS are found to be significantly higher than those for some of the other proteins studied so far, such as HP-36 and lysozyme. In particular, we find that more structured water molecules, with higher residence times (~300-500ps), are present near HFS than those near DFS. A significant slowing down is observed in the decay of associated rotational auto time correlation functions of O−H bond vector of water in the vicinity of HFS. The surface topography and the arrangement of amino acid residues work together to organize the water molecules in the hydration layer in order to provide them with a preferred orientation. HFS having a large polar solvent accessible surface area and a convex extensive nonpolar region, drives the surrounding water molecules to acquire predominantly a clathrate-like structure. In contrast, near the DFS, the surrounding water molecules acquire an inverted orientation owing to the flat curvature of hydrophobic surface and interrupted hydrophilic residual alignment. Water molecules near DFS are found to experience smaller free energy barrier heights separating them from the bulk water. We have followed escape trajectory of several such quasi-bound water molecules from both the surfaces and constructed free energy surfaces of these water molecules. These free energy surfaces reveal the differences between the two hydration layers.

# I. Introduction

The biologically active form of the protein Insulin is a monomer consisting of two chains. One chain (called A-chain) is 21 amino acids long and the other (called B-chain) is 30 amino acids long. These two chains are linked by two disulphide bridges at A7-B7 and A20-B19. The A-chain also has an intra-chain disulphide bond between A6 and A11. Insulin also exists, in its biologically inactive forms, as a dimer and a hexamer. [1] Although Insulin receptor signalling has evolved to facilitate Insulin binding only as a monomer to the Insulin receptor, it ensures that this important protein is stored in the body as a dimer or hexamer. [2] However, Insulin dimer, being relatively unstable, easily dissociates into monomers in blood circulation. The dimeric form in turn gets stabilized by the formation of the hexamer in the presence of zinc ions, during storage in the pancreatic β-cell. [3] So far several mutagenesis studies have generated different analogues of Insulin to tune its pharmacokinetic properties by mutating on different potent sites on hexamer and dimer forming surfaces. In this process Insulin often yields the analogue with reduced ability of forming any bio-aggregates. [3-5]

Water has a big role to play in Insulin activity. It is known from numerous medical reports that *dehydration tends to raise blood sugar* and can cause temporary resistance to Insulin causing "Diabetes mellitus". It is a known fact that water intake can significantly stimulate the function and dynamics of Insulin. It has been observed that plasma glucose decreased significantly in individuals after treatment with Insulin and the time of the maximum decrease (30 min) was synchronized with the beginning of water intake. [6] Hence there is a strong relationship between water and the function of Insulin that is yet to be understood at a molecular level.

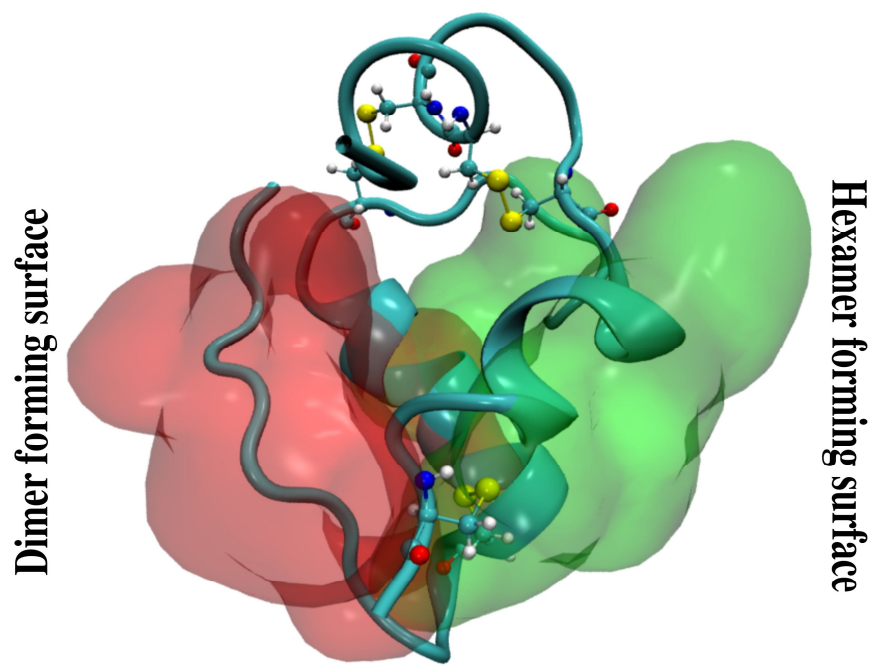

**Figure 1.** A ribbon based surface diagram showing two different surfaces of an Insulin molecule, with the red colour for dimer forming surface (DFS) and the green one for hexamer forming surface (HFS). One Insulin molecule is made of two chains: A chain and B-chain. These two chains are inter-connected though two disulphide bridges formed between A7-B7 and A20-B19. There is an intra-chain SS-bridge also formed between A7 and A11. The location of these three disulphide bonds those are essential for Insulin-receptor binding process and also in stabilizing the secondary structure are shown in the figure.

Despite extensive structural studies that have been done on Insulin, [7-10] relatively little research have been focused on the molecular dynamics of water molecules around this protein or on dynamics of this protein in water. In 1991 Mark et al studied the conformational flexibility of aqueous dimeric and monomeric Insulin. [11] In 1999, Rossky et al. performed molecular dynamics simulations of 2-Zn Insulin in water solvent to investigate the effect of vicinal polar or charged groups on hydrophobic hydration at a biomolecular surface. [12] In the same year, Chai and Jhon performed molecular dynamics study on interfacial hydration structure surrounding Insulin molecule at high pressure. [13] Since then the molecular dynamics simulations have become much more accurate and powerful. Moreover, the timescales in which studies can be performed have improved greatly since and a detailed dynamical study can now be carried out. While the water structure around Insulin has been studied by Chai et al. and Badger et al. [13, 14], to the best of our knowledge, no prior study exist that has evaluated the *dynamics of interfacial water* surrounding Insulin and the connecting role of such "biological water" with the structural morphology of the two distinct surfaces of Insulin.

It is known for a considerable amount of time that water in confined systems (like reverse micelles and nanotubes) and in biomolecular hydration layer exhibits properties different from bulk water. [15-21] Water in the protein hydration layer shows restricted motion in comparison to the free movement of molecules of bulk water. [22-27] Such water molecules are not only important for the thermodynamic stability of the proteins and DNA, but also play a central role in several biomolecular functions, such as intercalation, catalysis, recognition, etc. An example is provided by adenylate kinase (ADK) where water molecules help stabilizing a half-open-half-closed (HOHC) state that further facilitates substrate capture subsequent to product release. [28] Quasi-bound water molecules in the binding pocket of protein play a crucial role in noncovalent association of proteins and small drug compounds. The interaction between the binding site of protein and drug molecules depends upon the release of these bound water molecules. [29, 30]

The heterogeneity of an amphiphilic protein surface extends from largely hydrophobic to largely hydrophilic. While water molecules are usually found to form stable H-bonding network near hydrophilic residues through electrostatic interaction, [31-33] the vicinal hydrophobic patches often intervene. [34-38].

In a pioneering study, Rossky et al. have elucidated the effect of surface topography on the interfacial solvent structure in their numerous studies. They appear with the fact that the hydration of extended nonpolar planar surfaces that often appears to involve in structures is orientationally inverted in contrast to the clathrate-like hydration shell. While the clathrate-like structures predominate near convex surface patches, there is a dynamics equilibrium exists between clathrate-like and less-ordered or inverted structures in the hydration shell near flat surfaces. [39, 40] In a subsequent work they have explored the effect of vicinal polar and charged groups on hydrophobic hydration. In addition they have compared the hydration behavior around 2-Zn Insulin and Melittin dimeric surfaces. The dimer forming surfaces of Insulin is largely flat except with a slight curved exposure of Phe-25. Further the presence of vicinal polar residues causes fluctuation to such hydration structure. [41]

Dimer forming surface (DFS) and Hexamer forming surface (HFS) are both non-polar. [42] While DFS is flat and aromatic in nature, the other surface is more extensively nonpolar and convex. The DFS surface is buried upon Insulin dimer formation and HFS is buried when dimers assemble to form hexamer. The dimer forming surface (DFS) of Insulin is collectively hydrophobic (*effective hydrophobicity 1.4,* accroding to hydropathy scale) blended with a large number hydrophilic amino acids. DFS contains B8 Gly, B9 Ser, B12 Val, B13 Glu, B16 Tyr, B23 Gly, B24 Phe, B25 Phe, B26 Tyr and B27 Thr. Nevertheless the presence of residues like B12 Val, B24 Phe, B25 Phe being extensively hydrophobic in nature plays very important role in dimer link formation. The hexamer forming surface (HFS) of Insulin is also largely hydrophobic (,*effective hydrophobicity 10.9,* according to hydropathy scale) made of amino acids B1 Phe, B2 Val, B4 Gln, B13 Glu, B14 Ala, B17 Leu, B18 Val, B19 Cys, B20 Gly, A13 Leu, A14 Tyr and A17 Glu. This should explain, qualitatively, the weak stability of dimers that associate via the small hydrophilic surfaces and the enhanced stability of hexamers that aggregate through a large hydrophobic patch.

We organize the rest of the article as follows. In the next section we briefly describe the system studied (Insulin and water) and the simulation details. Section III contains the dynamical characterization of water near DFS and HFS of Insulin. Section IV involves the correlation between structure and dynamics. In this section we additionally show the movement of few interfacial water molecules along the escape direction. From the trajectory analysis, we obtain two-dimensional free energy surfaces of escape of those hydration layer water molecules towards bulk. Section V concludes with a brief summary of results.

## II. Structural Details of the Protein Studied and Simulation Setup

The monomer of Insulin is composed of two polypeptide chains, namely A-chain and B-chain. A chain involves 21 amino acid residues and B chain consists of 30 amino acid residues (see **Figure 1**). These two chains are inter-linked though a disulphide bridge formed berween A7-B7 and A20-B19. In addition A chains has an intra-chain SS-bridge formed between A7 and A11. All these three disulphide bonds are essential for the receptor binding activity of Insulin, as well as to preserve its secondary structural integrity. Consequently even among different species these three disulphide links along with certain amino acid sequences are highly conserved. [10] Such similarities in secondary structure rendering equivalent biological efficiency among different species have often been utilized in medication. For medical treatment pig Insulin is widely exploited in human patient.

To assess the dynamics of Insulin we performed molecular dynamics (MD) simulations of the protein in explicit water by using the Groningen Machine for Chemical Simulation (GROMACS Package). The simulation began with the crystal structure of the pig Insulin monomer. [43] The initial coordinate was collected from the Protein Data Bank (PDB-ID: 9INS). All atom topologies were generated with the help of pdb2gmx. We have applied the tricks of merging the two chains (A and B) and preserved the inter-chain disulphide linkage. We have treated the system with OPLS set of parameters available in GROMACS. Initially protein is cantered in a cubic box with length of 5.25Å. Then Insulin monomer was solvated with pre-equilibrated SPC/E water model using genbox. [44] A total of 4534 water molecules were added. After steepest descent energy minimization, each trajectory was propagated in a NVT ensemble and equilibrated for 2 ns. All the simulations in this study were done at 300 K and 1 bar pressure. The temperature was kept constant using the Nose–Hoover thermostat. [45, 46] It was followed by an NPT equilibration for 10 ns using the Parinello–Rahman barostat. [47] Finally, production runs were performed for each system in an NPT ensemble. Each simulation used a time-step of 2 fs. All the analyses were executed from the 20 ns trajectory. Periodic boundary conditions were applied and nonbonded force calculations employed with a grid system for neighbour searching. Neighbor list generation was performed after every 1 step. A cut-off radius of 1.0 nm was used both for neighbor list and van der Waal's interaction. To calculate the electrostatic interactions, we used PME [48] with a grid spacing of 0.12 nm and an interpolation order of 4. [49]

## III. Results : Dynamical Variation

### A. Reorientational Dynamics of Surface Water

Although a tentative picture of water arrangement around surfaces of different proteins have emerged in the recent years, no detailed study of water dynamics surrounding Insulin seems to exist. The reorientational dynamics of water molecules near the heterogeneous macromolecular surfaces (such as protein, DNA etc) is known to be significantly affected. [50] The reorientational motion of water can be evaluated by measuring the average time autocorrelation functions $C_i(t)$ of any bond vector $i$. The time correlation function (TCF) is defined as,

$$C_i(t) = \frac{\langle \hat{e}_i(t+\tau) \cdot \hat{e}_i(\tau) \rangle}{\langle \hat{e}_i(\tau) \cdot \hat{e}_i(\tau) \rangle} \tag{1}$$

Here $\hat{e}_i(t)$ is the unit vector of the corresponding bond at time, t.

To investigated the dynamical behaviour of water molecules both near HSF and DFS of Insulin, we evaluate the reorientational motion of the water molecules that are in the proximity to those surfaces (i.e., within 4.2 Å from the atoms of individual surfaces). The correlation functions were calculated by averaging over these water molecules only. In **Figure 2** we show the variation of $C_{O-H}(t)$ against time for water molecules near those surfaces. For comparison, we have also plot the relaxation for the bulk water. Note the significant slowing down of rotational motion of water molecules in the Insulin hydration layer when it is compared with that of the bulk water. Again, *it is clearly evident from the figure that the water molecules around DFS reorient noticeably faster than those around HFS*. This suggests relatively lower structuring of water layer around DFS. Exhibition of faster reorientational motion of water near DFS correlates well with the fact of less stable dimer formation than more stable hexamer.

Study of mean square deviation (MSD) of water molecules near the surfaces namely DFS and HFS also indicates that the DFS hydration layer shows comparatively faster dynamics than that around the HFS.

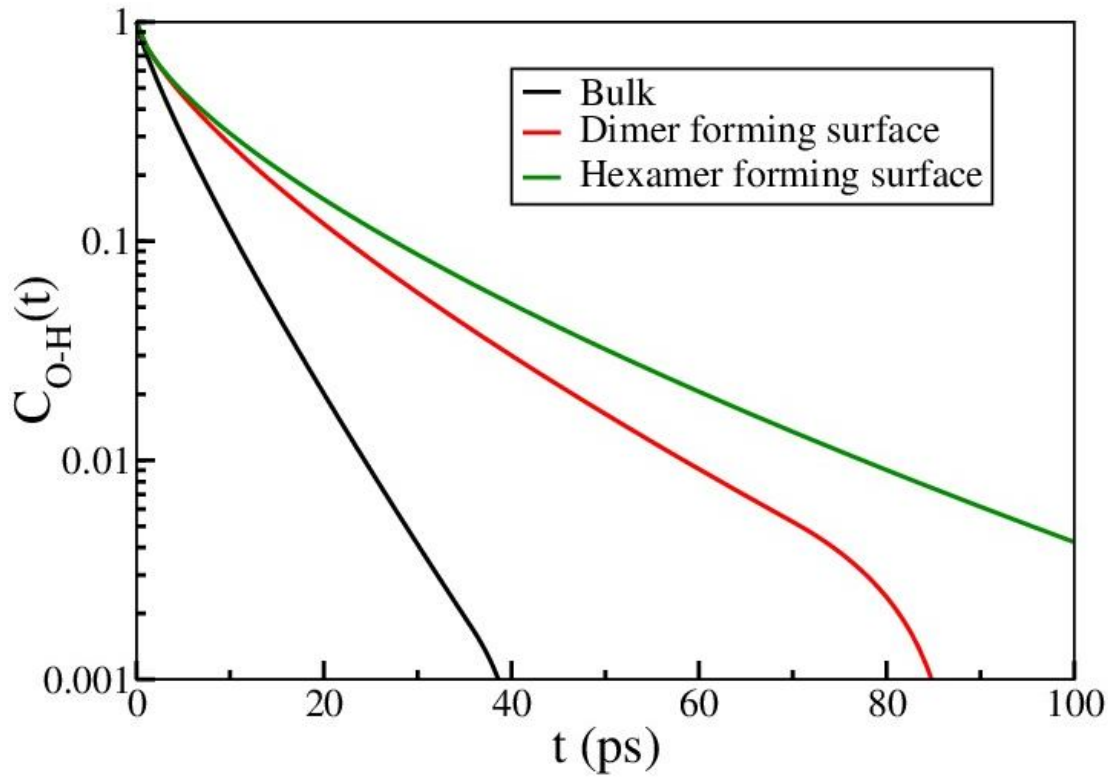

**Figure 2. Time evolution of reorientational time correlation function (TCF) of the O-H bond vector of water molecules near the two biologically significant surfaces of Insulin. The red curve shows the decay of TCF near the dimer forming surface (DFS) while the green curve shows the decay of TCF near the hexamer forming surface (HFS). The TCF for the water molecules present in the bulk (black curve) is also presented for comparison. Note the slow dynamics in relaxation of TCFs both near DFS and HSF. However, the emergence of long time tail for water TCF near hexamer forming surface (HFS) is noticeable.**

**Table 1: Stretched exponential fitting parameters for the reorientational time correlation functions of interfacial water molecules around the DFS and HFS of Insulin also in comparison with bulk.**

|  | $\tau$ | $\beta$ | $\langle\tau\rangle$ |
|---|---|---|---|
| **Bulk** | 3.95 | 0.84 | 4.32 |
| **DFS** | 7.0489 | 0.71 | 8.64 |
| **HFS** | 7.879 | 0.6347 | 10.41 |

Furthermore, to understand the observed time correlation functions in a more quantitative way, we have estimated the time scales associated with those quasi-bound water molecules located next to both surfaces. We notice that even though the water molecules around DFS reorient more quickly compared to those around HFS, all the curves show slower decay at longer times. Such long-time decay cannot be described by a single-exponential law. Here we observe the computed TCFs are best fitted by stretched exponential functions of time with stretching exponents (β) in the range of 0.6-0.8 that provides reasonable fits of the data. The expression used for the best fit is as follows:

$$C_{O-H}(t) = e^{-(t/\tau)^\beta} \tag{2}$$

The parameters for the best fit are shown in **Table 1**.

It is clear that the water molecules around HFS exhibit significantly slower dynamics with a long time component. In the recent past, similar slow decay has also been observed in a number of related studies for protein HP-36 and enterotoxin. As discussed above, the existence of such a long time component arises because of particular water molecules that are "quasi-bound" to specific residues on protein surfaces. The present result and estimated time scales clearly suggest that even though the rotational motion of water at the interface of a protein is much slower compared to that of bulk water, significant difference in water motion might arise due to distinct surface topography. To obtain a microscopic understanding of such diverse dynamical behaviour and it likely influence on the binding activity of the protein, it would be interesting to study the hydrogen bond forming affinity of water molecules with the adjacent protein residues.

**B. Hydrogen Bond Lifetime Kinetics in the Two Layers**

Depending on the number and nature of hydrogen bonds (H-bond) that these water molecules make with the charged/polar amino acid residues on the protein surface, we can divide them broadly into two classes: (i) interfacial quasi-bound water (IQBW) and (ii) interfacial free water. [51] These interfacial free water molecules do not form any hydrogen bonds with the protein residues whereas interfacial quasi-bound water molecules might exist in either singly or doubly hydrogen bonded form. Interfacial free water molecules of course form hydrogen bonds with neighbouring water molecules and experience van der Waals type interactions with protein atoms if they are within the range of any sort of interaction potential. [52, 53] One can use either a geometric or an energetic criterion to define a hydrogen bond. In the present work, we have applied solely the geometric criterion to define a hydrogen bond. [50]

The structural relaxation of water molecules in terms of hydrogen bonds can be expressed as,

$$C_{HB}(t) = \frac{\langle h(t+\tau)h(\tau) \rangle}{\langle h \rangle} \qquad (3)$$

According to the definition the hydrogen bond population variable, h(t) is unity when a particular pair of protein−water or water−water sites is hydrogen-bonded at time t and is zero otherwise. The angular brackets denote averaging over all protein−water hydrogen bonds and over initial time τ. Here we have not presented any water-water hydrogen bond dynamics. The correlation function $C_{HB}(t)$ allows the re-formation of a bond that was broken at some intermediate time. In fact it allows recrossing the barrier separating the hydrogen bonded and non-bonded states. Thus, the relaxation of $C_{HB}(t)$ offers the information about the structural relaxation of a particular hydrogen bond. The computed hydrogen bond TCFs are best fitted by stretched exponential functions of time with stretching exponents (β) in the range of 0.6-0.8 that provides reasonable fits of the data. The parameters for best fit are shown in **Table 2**. In the present case we have evaluated hydrogen bond time correlation function for water molecules that specifically are close to either DFS or HFS (see **Figure 3** for $C_{HB}(t)$). Both reorientational and hydrogen bond time correlation function calculation were carried out from the simulation trajectories with the time resolution of 2 fs.

The most interesting observation is the significant difference in the relaxation behaviour of hydrogen bond dynamics in DFS-water H-bonds and HFS-water H-bonds. Similar as above figure here also the structural relaxation of the protein−water hydrogen bonds is much slower for HFS than that of DFS. These results also correlate well with the biological functionality of the protein because most of the hydrophobic residues of Insulin are congregated in HFS to form stable hexamer.

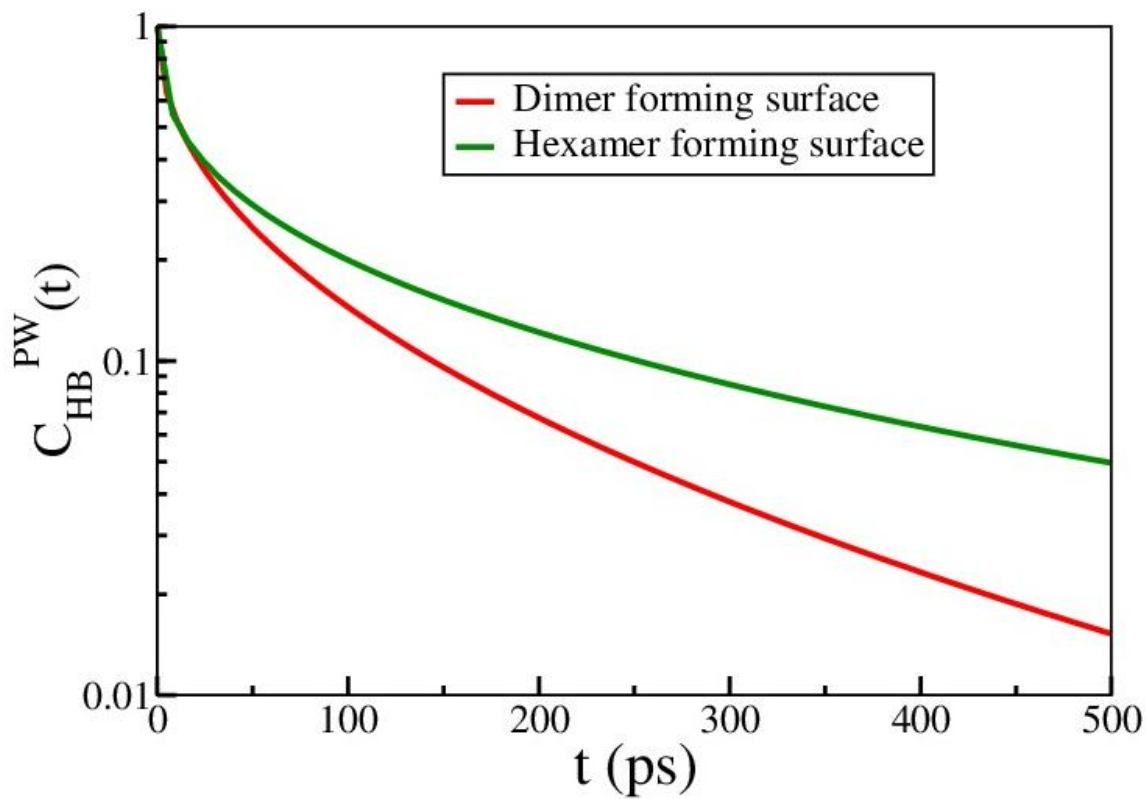

**Figure 3.** Hydrogen bond lifetime dynamics of water molecules near the two surfaces: DSF (red) and HFS (green). Note the slower hydrogen bond relaxation behaviour near HFS.

**Table 2:** Stretched exponential fitting parameters for the hydrogen bond time correlation functions of interfacial water molecules around the DFS and HFS of Insulin.

|  | $\tau$ | $\beta$ | $\langle \tau \rangle$ |
|---|---|---|---|
| **DFS** | 25.299 | 0.4796 | 50.45 |
| **HFS** | 29.3119 | 0.388 | 71.919 |

## C. Residence Time Distribution of Hydration Water

The slow dynamical behaviour of water near HFS and DFS provides the signature of the presence of a number of quasi-bound water molecules that are motionally restricted at the Insulin surfaces. To distinguish the propensity and survival of the existence of such quasi-bound water molecules in the hydration layer, we evaluate their residence time distribution, shown in **Figure 4**. The residence time distributions, for both the hexamer and dimer forming surface, peak at around 150 Ps which is unexpectedly high. This seems to indicate the water layer structured to a certain extent. Such a long residence time has not been observed in other similar studies such as those with HP-36. [51]

A curious phenomenon is observed for the water molecules around the hexamer forming surface. Some water molecules have even a residence time extending to 700ps which is unusually high. These water molecules are found to be mostly either singly, doubly or triply hydrogen bonded to the protein residues. The population of such water molecules is in never large but non-negligible near the HFS. But a much larger population in the more mobile (low residence time) region ( less than 150 ps) may suppress the contribution from those "slow" interfacial water molecules which are less in number but more strongly bound to the protein surface. To separate out those strongly bound water molecules we have zoomed in the time frame towards the long time beyond 100ps as shown in the **inset of Figure 4**. In this plot a clear appearance of the long time tail of residence time near HFS illustrates that there exist a fraction of water molecules with residence time in the range of 400-700 ps. The number of such quasi-bound water molecules with long survival time is indeed very less.

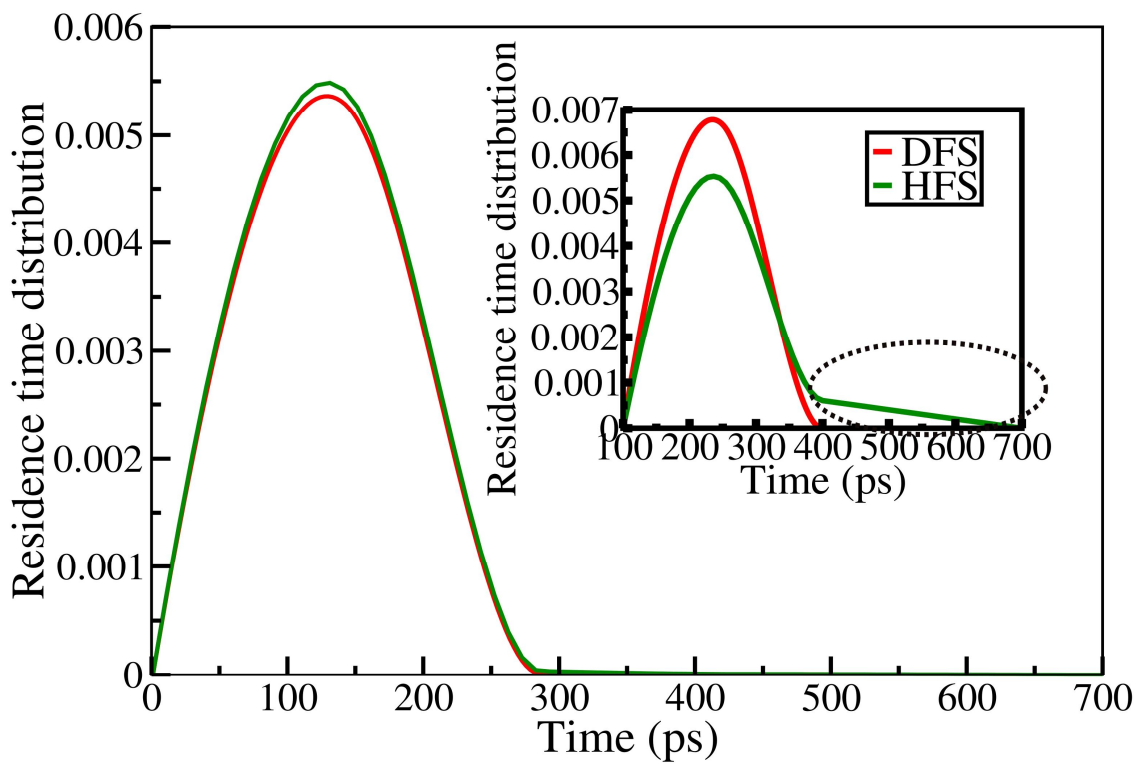

**Figure 4.** Residence time distribution near the DFS and HFS surfaces. The color code is the same as in the previous figures. The residence time distributions of water molecules that stay in the hydration layer more than 100 ps are highlighted in the inset of figure. Note the bimodal nature in the plots of the residence time distribution near HFS.

## IV. Correlation between Structure and Dynamics

### A. RMSD

To understand the dynamical coupling between the conformational fluctuation of protein and surrounding water molecules, we have monitored the root-mean-square deviation (RMSD) of position of all non-hydrogen atoms involved in the whole protein, particularly those involved in forming the DFS and the HFS. This is shown in **Figure 5**. RMSD often provides key information about the side chain mobility of the residues which can influence and also be influenced by the hydration layer dynamics. It is interesting to note that highly hydrophobic hexamer forming surface has a lower root mean square deviation, on the whole, which implies that it is more rigid than the dimer forming surface or than the whole protein. Such rigidity assists in building up a stable hydration layer in the HFS premises. There are several reports suggesting that protein dynamics is slaved by the water dynamics. The present result indeed is a good example of the same.

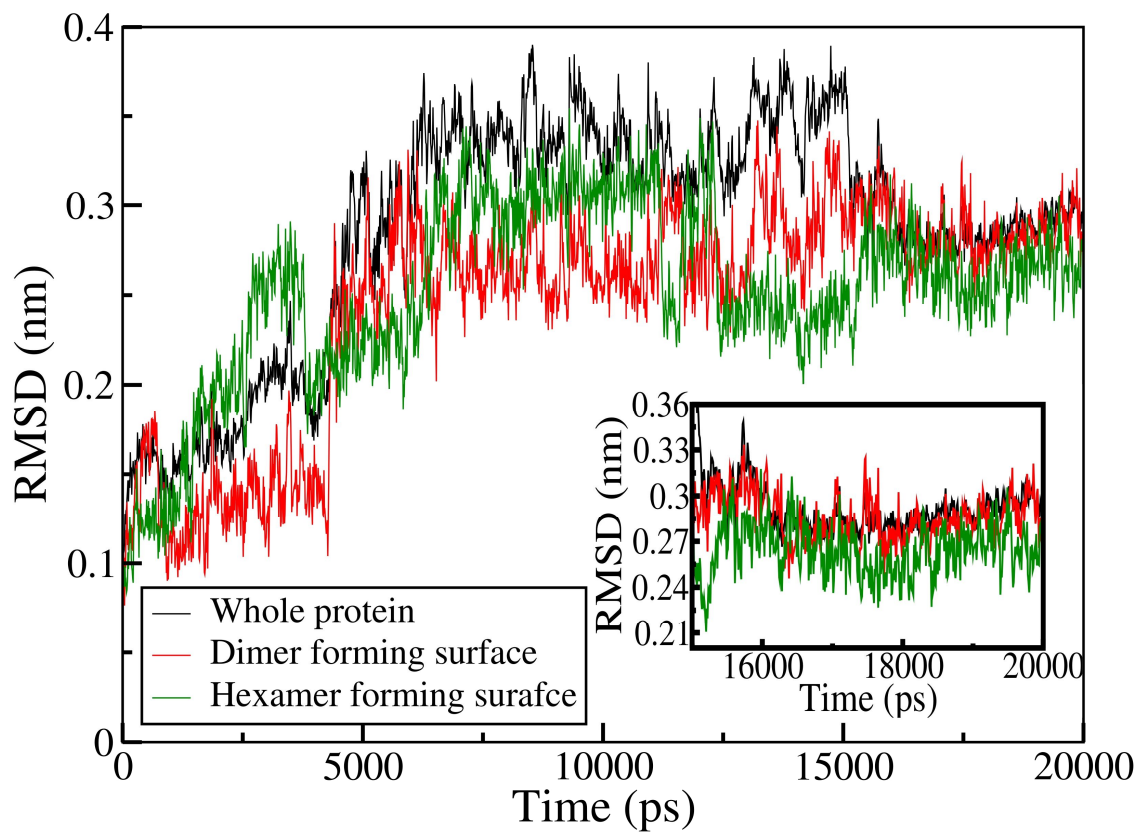

**Figure 5. Time depependence of the root mean square deviation (RMSD) of the whole protein (shown in black) of the dimer forming surface, DFS (shown in red) and the hexamer forming surface, HFS (shown in green). The equilibrium RMSDs for the two surfaces are highlighted from 15000ps to 20000ps in the inset which shows remarkable lowering of the RMSD of HFS. The higher RMSD values for DFS residues indicate a more flexible environment around that specific region.**

## B. Correlation Of Water Dynamics with Effective Hydropathy Index and SASA

The hexamer forming surface consists mostly of highly or moderately hydrophobic residues : Phe( B1), Val B2, Ala(B14), Leu(A13 andB17), Val(B18), Cys(B19) . Others are hydrophilic at neutral pH:  Gln (B4) and Glu (B13 and B17), Gly (B20) and Tyr (A14).  The hexamer forming surface has a higher relatively non-polar significantly exposed and extended convex hydrophobic solvent accessible surface area (see **Table 3**). In contrast, the dimer forming surface involves less hydrophobic residues namely Val (B12) and Phe (B24 and 25), Leu (A13 and B17) and more hydrophilic residues, such as, Ser (B9), Glu (B13), Tyr (B16 and 26) and Thr (B27) along with Gly at B8 and B23 position. Thus a higher polar solvent accessible surface area of DFS is expected. Following the hydropathy scale index we have calculated an effective hydropathy index considering all the involved residues along both surfaces. It provides an intuitive idea to understand and explain the events of a specific surface. This hydropathy estimate shows that while DFS has an overall hydrophilic index -8.4, HFS has overall hydrophilic index -12.2. More interestingly, while DFS has overall hydrophobic index 9.8, HFS has overall hydrophobic index 23.2. Thus HFS with elevated hydrophilic and hydrophobic character might be responsible for observed dynamical retardation.

Other than the hydropathy estimates the relative exposure of the polar probe residues in different surfaces might play a role which determines the essential part of protein-water interaction causing such slow dynamics. To estimate the exposure of the polar probe we have calculated relative polar solvent accessible surface area averaged over 20ns trajectory (shown in **Table3**). We find that the solvent accessibility to the polar probe is low near DFS (28.8%) compared to that of HFS (35.9%). Despite the presence of extended hydrophobic surface we have observed that HFS is designed in such a way that polar residues are next to another polar neighbour that assists to bind the water molecules near HFS.

Table 3. Estimation of effective hydropathy index in terms of overall hydrophilicity and overall hydrophobicity following the hydropathy scale. In comparison the evaluation of relative polar and nonpolar SASA (solvent accessible surface area) of DFS and HFS is also indicated in the table.

| Surface | Overall Hydrophilicity | Overall Hydrophobicity | Polar SASA | Non-polar SASA |
|---|---|---|---|---|
| DFS | -8.4 | 9.8 | 28.8% | 71% |
| HFS | -12.2 | 23.1 | 35.9% | 63.1% |

## C. Role of Protein-Water Electrostatic Interaction Energy

The time evolution of protein-water electrostatic interaction energy per residue for both DFS and HFS are shown in **Figure 6.** The figure shows that the quasi-bound water molecules near HFS are displaying higher stability with more negative interaction potential. Such quasi-bound water molecules are stabilized by about 20 kJ/mol energy compared to DFS. It is mostly arises due to doubly hydrogen bond formation and electrostatic interaction with the protein residues. However, these quasi-bound water molecules have lower entropy which is found to play an important role in determining their overall stability.

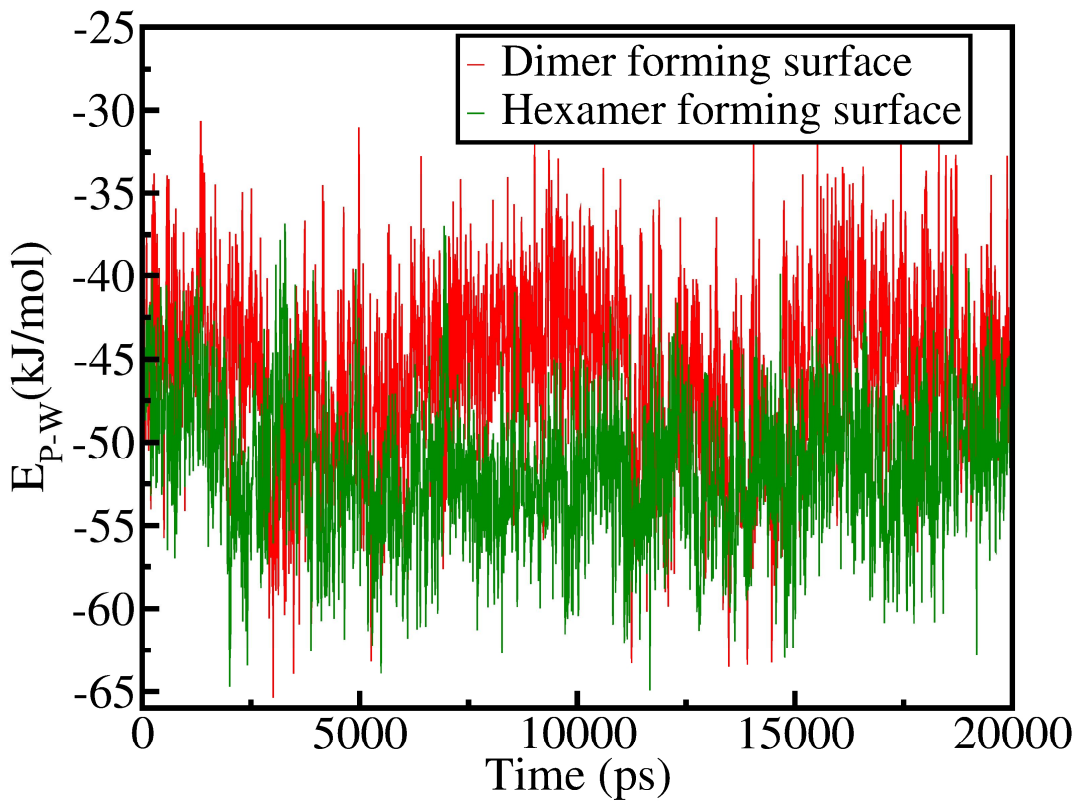

**Figure 6.** Time progression of protein-water electrostatic interaction energy per residue. Comparison between the relative electrostatic interaction energy for DFS and HFS clearly shows the higher stability of HFS hydration layer.

## D. Quasi-bound Water Crowding near DFS and HFS

We detect a few motionally restricted water molecules located near HFS that are doubly hydrogen bonded to the protein residues. When we extracted several snapshots from simulation trajectory we observed that a large number of water molecules with high survival time are crowded near HFS. In **Figure 7** we show such a cluster of water molecules that are captured in one such snap. In comparison to HFS, less number of water assemblies is found near DFS (see **Figure 7**). The higher residence time of the hexamer forming surface (HFS) may be attributed to the presence of an exposed hydrophilic residues (B4 Gln, B13 Glu, A14

Tyr and A17 Glu), in combination with the presence of a largely hydrophobic surface which restricts the options to avail that unfavourable surface. The encaged water molecules thus revolve around that hydrophobic area for a long time and thus account for the unusually high residence time in the hydration layer. On the contrary, the hydrophilic groups that are involved in DFS are mostly located between two hydrophobic shells of DFS and HFS. Such hydrophilic residues those are located in the intersection of two extended hydrophobic region themselves exhibit large fluctuations in position rendering the location unstable (or, unsuitable) for water.

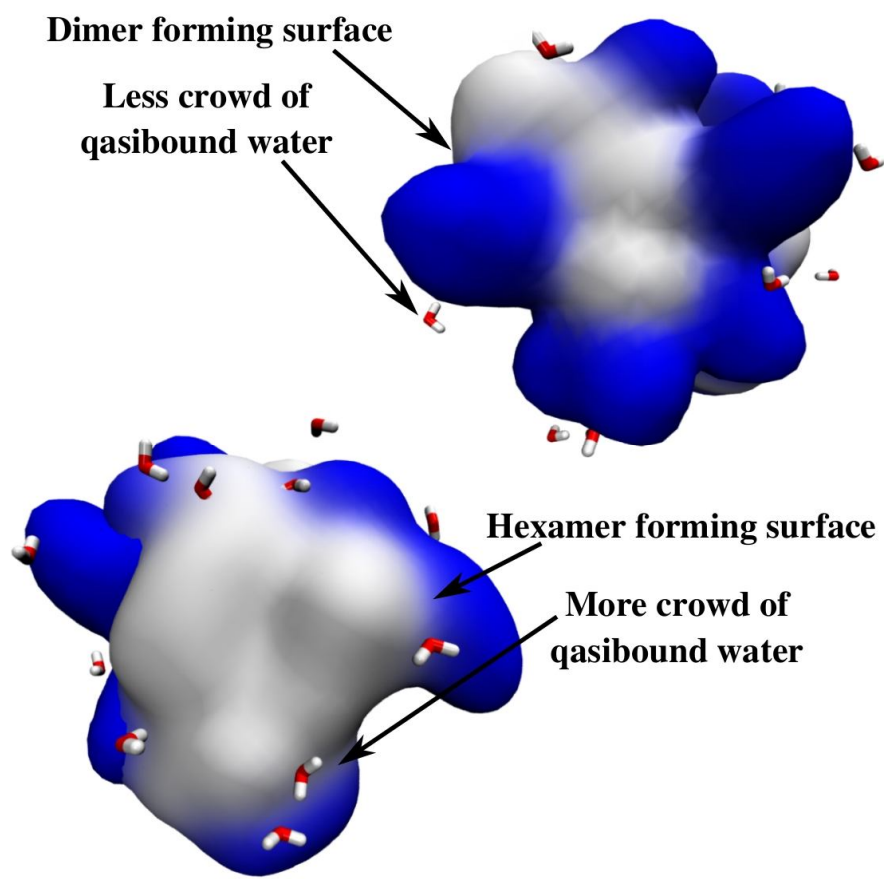

**Figure 7. Representative snaps extracted from the simulation trajectory, showing the location of several water molecules adjacent to the surfaces of DFS (upper) and HFS (lower), respectively.**

**The number of quasi-bound water molecules attached to the dimer forming surface is less than the number bound to the hexamer forming surface. Note that quasi-bound water channel around the extensive hydrophobic surface of HFS.**

Although the hydrophilic residues and their spatial arrangement induce the water molecules to stay in the hydration layer through electrostatic interaction, the aversion of extended hydrophobic patch is also believed to play a role in their stay in the hydration shell. To address the surface topographical dependence of hydrophobic hydration (by the proximal water molecules) we have looked into the water orientation along the protein surface by following a scheme invented by Rossky and co-workers. [40] In this analysis we consider the resultant water dipole vector of water pointing tetrahedrally outward from the central oxygen atom as it can form four hydrogen bonds. Now we measure the angle ($\theta$) between the vector of this resultant dipole and tangential vector associated with the protein surface normal and oxygen of water. See **Figure 8** for an illustration. The probabilistic distribution of cosine angle reflects the structure of the hydration shell. We find that the distribution maximizes sharply near -1 and 1 that are conventionally belonging to clathrate-like hydration shell and inverted hydration shell respectively. Additionally we observe two broad maxima appear over the range of –0.33 to -0.42 near inverted region and on the other side over the range of 0.33 to 0.50 near clathrate-like hydration shell. These values might correspond to the other connected hydrogen bonds that are spatially form tetrahedral arrangements. In a previous study Rossky et al. has detected similar values in such distribution. [40]

The investigation of water orientation correlates well with the topographical construct of the corresponding hydrophobic surface of DFS and HFS. In DFS the residues are arranged in a flat surface. [43] Although in DFS, Phe (B25) being more convex and exposed prefers a clathrate-like hydration arrangement, Phe (B24), Val (B12) residues prefer an inverted hydration shell due to the influence of adjacent polar residues such as Arg (22) and Glu (B13) respectively. In HFS, while a large number of residues are exposed, such as Phe( B1), Val (B2), Leu(A13 andB17), Val(B18), two residues, Ala(B14) and Cys(B19) are found to be buried. However a large hydrophilic surface exposure and highly convex restricted hydrophobic area consent to the hydration shell to slightly shift towards a more clathrate-like distribution.

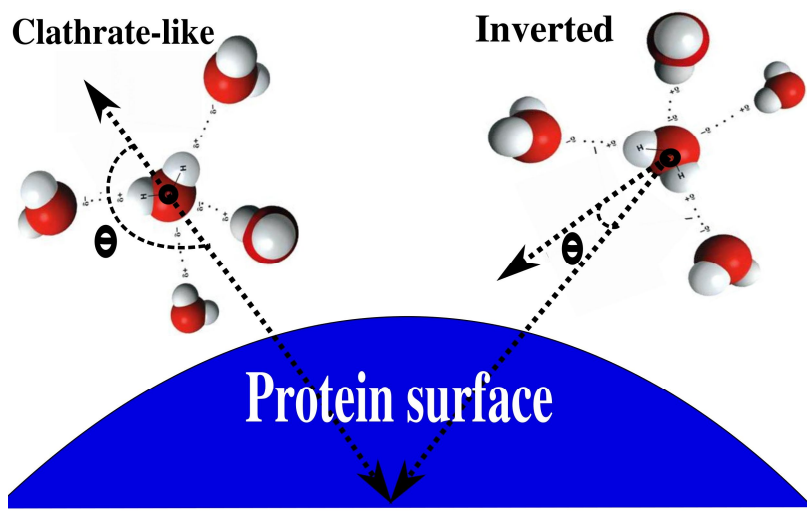

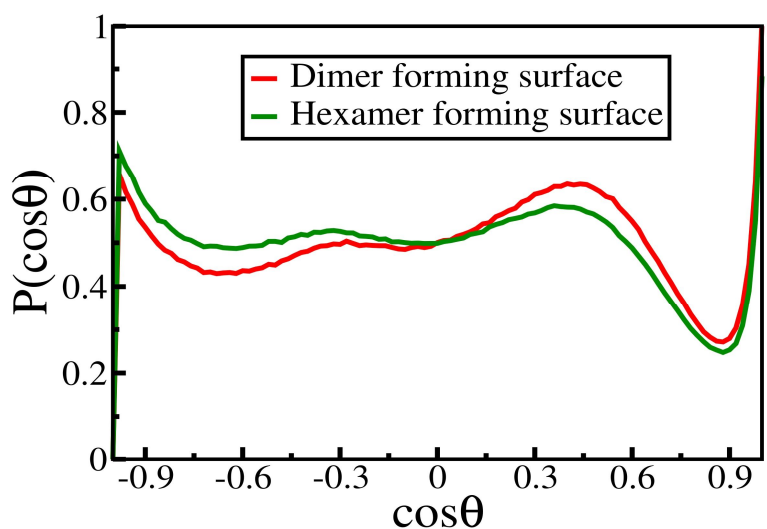

**Figure 8:** Orientational distribution of water molecules in the hydration shell relative to the normal of the protein surface. Upper panel provides a pictorial description of the angle θ, between the vector of water dipole and the protein interface. θ categorizes water molecules into one of the two forms: (i) either clathrate-like arrangement if the cosθ value is near -1 or (ii) inverted arrangement if cosθ value is 1. A broad maximum appears over the range of –0.33 to -0.42 which corresponds to the orientation of those water molecule that are tetrahedrally connected outward with an inverted water molecule. On the other hand, over the range of 0.33 to 0.50, another broad maximum appear which corresponds to the orientation of those water molecule that are tetrahedrally linked inward with a clathrate-like water molecule in the hydration shell. [40]

### E. Details of Water Motion in Insulin Hydration Layer

We have followed the movement of several strongly hydrogen bonded water molecule near DFS and HFS (see **Figure 9**). From DFS we tag one water molecule which is often doubly hydrogen bonded to the OG1 atom and N atom of Thr (B27) and substantially form a third hydrogen bond with the OH group of Tyr (B26). (See **Figure 9(a)**) It stays around 187.66 ps in the 1$^{st}$ hydration layer. For HFS we follow the movement of a water molecule which is doubly hydrogen bonded to OE1 atom of Gln (B4) and N atom of ASN (B3) (see **Figure 9(c)**). Water motion on the protein surface can be decomposed into two types of the movement, that in the tangent (with respect to the protein surface) direction and the other along the normal direction. We investigate both the in and out motion of the interfacial quasi-bound water molecules along the escape direction (Z-axis) and the lateral or parallel motion along the protein interface (here represented along X-axis). We find from the trajectories that the water molecules move mostly along the Z-direction to escape from its hydration layer, and hence the Z-direction is designated as the escape direction. Both of the three-dimensional trajectories are followed during its stay within 7Å (corresponding to the 2nd minima of the radial distribution function of oxygen atom of water) from the protein surface. The dense part of the trajectory shown in **Figure 9(b)** indicates the restricted movement of quasi-bound water molecule within the hydration layer of DFS. During this time interval, the quasi-bound water molecule visits some stable regions of protein where it carries out the dangling motion. Afterward this water molecule moves laterally as well as in the perpendicular direction over a distance corresponding to the escape from the hydration layer. The trajectory of interfacial quasi-bound water near HFS is somewhat different from the quasi-bound water molecules near DFS (see **Figure 9(d)**). The highly restricted motion during its residence time signifies higher stability due to the strong interaction with the protein residues.

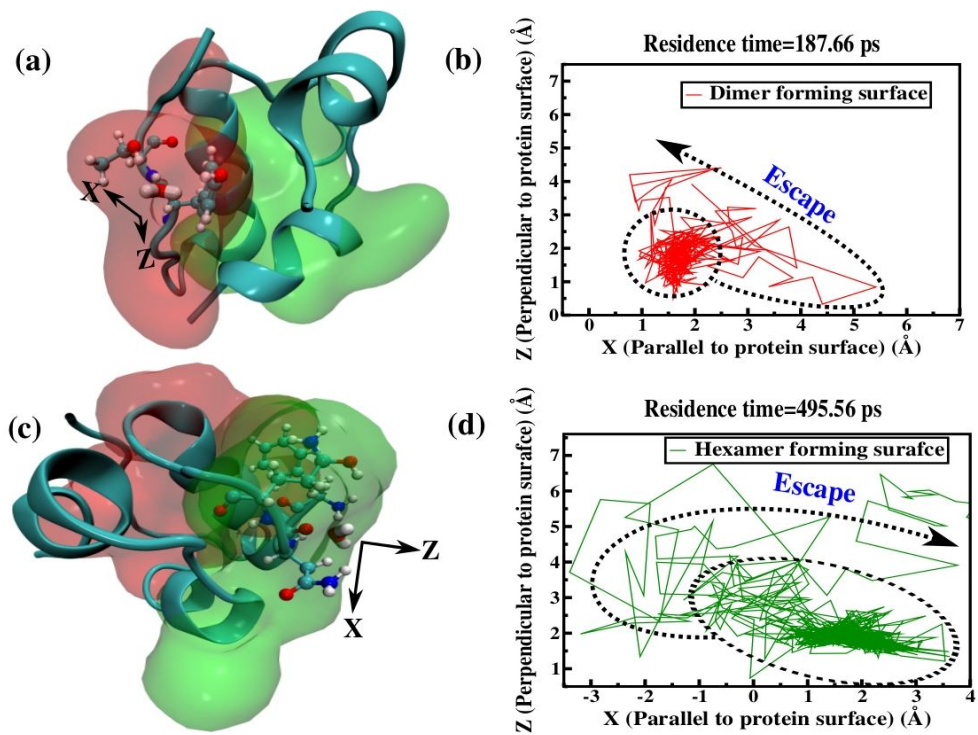

**Figure 9. X and Z-coordinates of (a) quasi-bound water molecule near DFS, and (c) quasi-bound water molecule near HFS. The Z-coordinate represents the direction of escape of the**

**water molecule from protein interface to the bulk region where the X-axis corresponds to the protein interface. The projection of the trajectory of the water molecule is shown along the escape direction (Z) and along the protein surface (X) for quasi-bound water molecule near (b) DFS and (d) near HFS. The trajectories are collected during their stay within 7 Å. Note the escape along Z-direction for both of the cases.**

## F. Free Energy Surface from Trajectory Analysis

We evaluated free energy surfaces of the tagged interfacial water molecules from the histogram of the above trajectories. The contour map of the free energy surface for strongly hydrogen bonded water molecule shows the presence of two distinct minima along the protein interfacial plane of DFS bearing its doubly hydrogen bonded character. The deep minimum in **Figure 10(a)** probably arises from the breaking of one strong hydrogen bond and the comparatively shallow minimum appears due to loss of another hydrogen bond formed with the protein residues. The existence of two minima is a signature of the presence of more than one hydrogen bonding forming centre due to the presence of a number of hydrophilic residue. However, along the Z-axis, (which is characterized as the direction of escape), its escape to the bulk is evident from the plot. On the other hand, in the case of quasi-bound water molecules near HFS, we find only one deep minimum (see **Figure 10(b)**) which is due to the presence of one very strong hydrogen bond interaction forming with the hydrophilic residue in HFS. This bears the signature of a lower probability to find other such hydrogen bonding interaction centre.

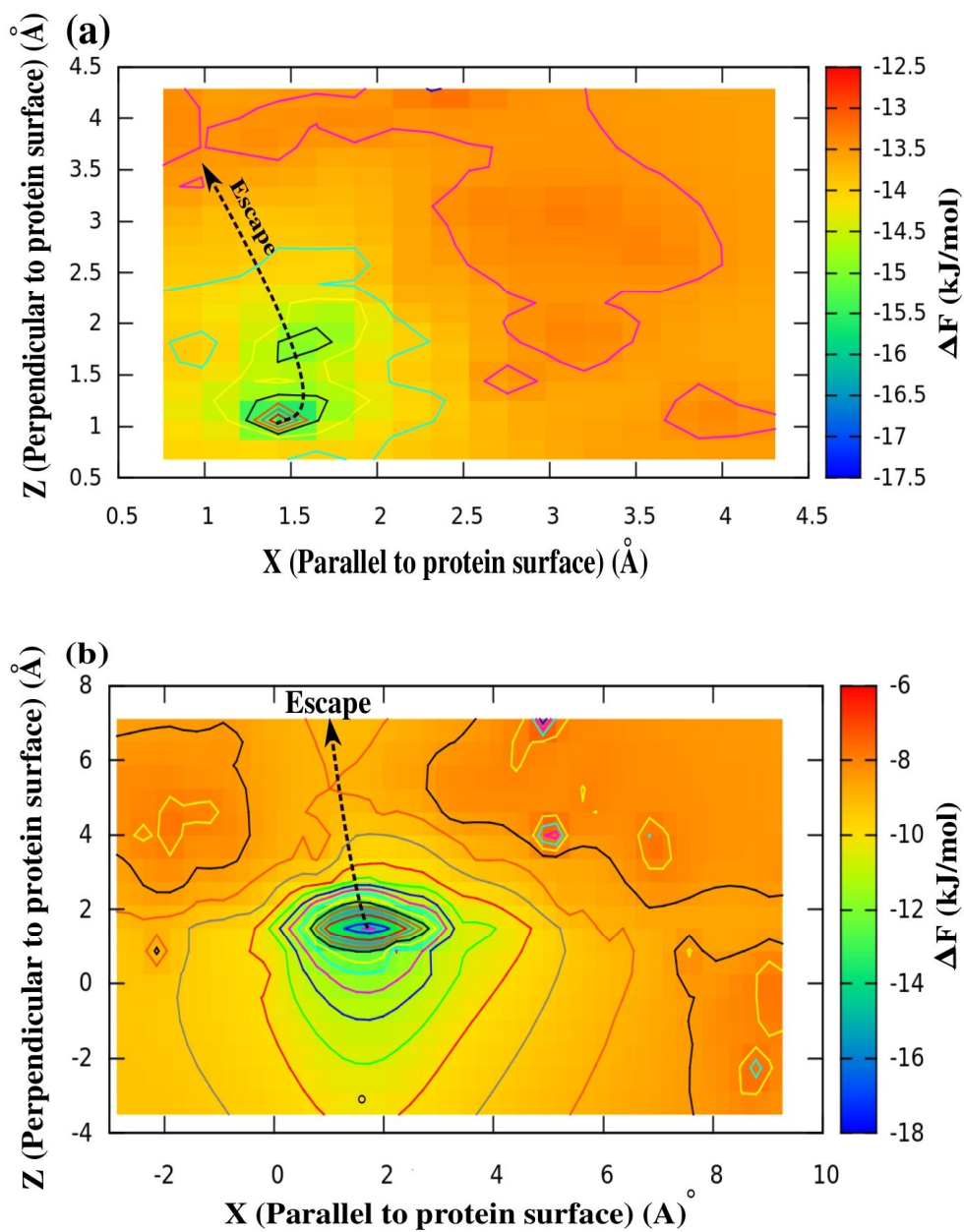

**Figure 10.** Contour map of free energy surface of quasi-bound water molecules near (a) dimer and (b) hexamer forming surfaces depicting the saddle region of each, for escape direction from the surface. Note the presence of two minima in (a) for water molecule in strongly H-bonded to DFS as shown in figure 9(a). Here the presence of two minima corresponds to two hydrogen bond breaking events. In (b) note the existence of a single deep minimum indicates only one strong hydrogen bond rapture of the quasi-bound water molecules as shown in Figure 9(b). For both cases, the escape along the Z-direction is evident from the contour. The colour code of the free energy landscape has been so chosen that the closely spaced regions can be distinguished clearly.

## V. Conclusion

Given the paramount importance of water in the biological function of Insulin, it is interesting to find the substantial differences in the solvation characteristics of the two surfaces of an Insulin molecule. The observed differences could have important consequences in dimerization and aggregation of these protein molecules. We have obtained several potentially important results. First, the residence times of water molecules in the protein hydration layer for both the DFS and HFS are found to be significantly higher than those known for some other proteins that have been studied so far, such as HP-36 and lysozyme. Crystallographic studies have also identified the presence of such structurally ordered water molecule in Insulin hydration layer. [54] In particular, we find that more structured water molecules, with higher residence times (~300-500ps), are present near HFS than those near DFS. Second, dynamical characterization reveals that kinetics of the hydrogen bonds formed between the residues of DFS and the interfacial water molecules are faster than that formed between water and the HFS. Third, a significant slowing down is observed in the decay of associated rotational auto time correlation functions of O−H bond vector of water in the vicinity of HFS.

The dynamical behaviour of interfacial water molecules is complex and often defies any generalization. There are several views on the complex movement of hydration layer water molecules depending on protein surface topography. In recent past, the hydration structure of human lysozyme was analyzed by using molecular dynamics simulations by Jana et al and several similar reports also exist associated with other proteins. [32, 24, 50] These studies found both fast water motion and stable hydrogen bonding (H-bond) network near hydrophilic patches, reflecting the role of electrostatic interaction between the polar amino acid residues. Several studies have also revealed structural ordering of water molecules near hydrophobic surfaces.

In the present case the reason behind distinct dynamical behaviour cannot be attributed solely to the hydrophilic-hydrophobic proportion but due attention should be given also to their length scale and relative arrangement of the groups. While the HFS, so designed, has a large hydrophilic exposure blended with an extensively restricted hydrophobic convex area. Thus the water molecules those are strongly hydrogen bonded to the hydrophilic residues of HFS become more constrained and structured. The hydrophilic residual alignment also providing cooperativity to the adjacent polar group assist to build a stable water channel surrounding the reviled hydrophobic area. The convex nonpolar residues all along restricting the water motion direct them to orient in a clathrate-like arrangement. The dynamical slowing down in HFS is attributed to such ordering in the hydration structure.

However, in the case of DFS, the hydrophilic residues are often interrupted by the intervening hydrophobic moieties. Such interventions lead a fluctuating hydration shell around DFS. In addition a large portion of DFS hydrophobic surface is flat. Such topography directs the vicinal water molecule to orient in an inverted arrangement. Yet the highly fluctuating protein dynamics near DFS leads the hydration structure to be less structured and thus we obtain a relatively fast water dynamics in DFS hydration shell.

Water molecules near DFS experience smaller free energy barrier height that separates them from the bulk water. We have traced the trajectory of several such quasi-bound water molecules from both the surfaces towards their escape into the bulk. The "slow" water molecules may play an important role in stabilizing hexamer forming surface of the protein and thus, perhaps, assist in the formation of stable bio-assembly. This aspect deserves further study.

## Acknowledgement

This work was supported by grants from DST, India. We thank Prof. B. Bagchi for many useful discussions.